\documentclass{kluwer}    % Specifies the document style.
\usepackage{graphicx}
\begin{document}
\begin{article}
\begin{opening}
\title{Preliminary analysis of photometric variations of central star of planetary nebula
Sh 2-71}
\author{Zden\v{e}k \surname{Mikul\'{a}\v{s}ek}$^{1,\,3)}$, Lubo\v{s} \surname{Kohoutek}$^{2)}$,
Miloslav \surname{Zejda}$^{3)}$,
Ond\v{r}ej~\surname{Pejcha}$^{3)}$}

\institute{$^{1)}$\,Institute of Theoretical Physics and
Astrophysics, Masaryk University, Kotl\'{a}\v{r}sk\'{a}~2,
611~37~Brno, Czech Republic
\newline $^{2)}$\,Hamburg Observatory,
Hamburg, Germany
\newline $^{3)}$\,N. Copernicus Observatory and
Planetarium, Krav\'{\i} hora 2, 616~00~Brno, Czech~Republic}

\runningauthor{Z. Mikul\'{a}\v{s}ek et al.}
\runningtitle{Preliminary analysis of photometric variations of
central star of Sh 2-71}
\date{May 15, 2004}

\begin{abstract}
We confirmed the presence of regular \emph{UBV}$\!(\!RI)_{\rm{C}}$
light variations of the object in the centre of planetary nebula
Sh 2-71, the improved period of them being $P$=(68.132$\pm$0.005)
days. The shapes and amplitudes of light curves in particular
colours are briefly discussed.
\end{abstract}
\keywords{\emph{UBV}$\!(\!RI)_{\rm{C}}$ photometry, planetary
nebula, Sh 2-71, period, PCA}

\end{opening}

\section{Introduction}
Nebula Sh 2-71 (= PK 036-01 1) was discovered by
\inlinecite{Minkowski1946} who classified it as a diffuse or
peculiar nebulosity. \inlinecite{Sharpless1959} included this
object in his catalogue of H II regions and considered it as a
possible planetary nebula.

\inlinecite{Kohoutek1979} reported light variation ($>0.7$ mag) of
the star in the centre of planetary nebula Sh 2-71 and
preliminarily classified the central star as B8 spectral type.
\citeauthor{Sabbadinetal1985} (\citeyear{Sabbadinetal1985},
\citeyear{Sabbadinetal1987}) showed the nebula as a high
excitation PN, having strong stratification effect. The very high
effective temperature of the ionizing source of 129~000~K was
determined by \inlinecite{Preitemartinez1989} and confirmed by
(\opencite{Kaleretal1990} and \opencite{Bohigas2001}).
\inlinecite{Feibelman1999} showed a variable Mg~II~$\lambda$~2800
emission line superposed on a variable stellar continuum.
\inlinecite{Jurcsik1993} revealed the periodical light variations
of the central star  with the period of 68.064 d, variations of
the colours in \emph{UBV}$\!(\!RI)_{\rm{C}}$ system she found
insignificant.

\section{Observations}

We have processed a large set of 3370 observations from three
sources (see the Table I):

\noindent 1. Partly unpublished \emph{UBV} observations made by
Kohoutek at Wise Observatory at Mitzpe Ramon, Israel, in
1977\,-\,79, at the European Southern Observatory at La Silla,
Chile and at the Hamburg-Bergedorf Observatory, Germany, in 1979.

\noindent 2. Unpublished \emph{UBV}$\!(\!RI)_{\rm{C}}$
observations made by \inlinecite{Jurcsik2003} at Konkoly
Observatory in 1990\,-\,93.

\noindent 3. Unpublished \emph{VRI} CCD photometry made by M.
Zejda, P. H\'{a}jek, O. Pejcha, J. \v{S}af\'{a}\v{r}, P. Sobotka
 at Brno in 1999\,-\,2002 and Vy\v{s}kov Observatories in 2001\,-\,2002.

Stars \emph{a} and \emph{b} (Kohoutek 1979) were measured as
comparison stars and HD 175\,544 (B3 V) as photometric standard
\cite{Menziesetal1991}.

\begin{table} %
\begin{tabular}{lrrrrrrr}
\hline
Author  & \emph{time interval~~~}  & $U~$ &  $B~$  &  $V~$  &  $R_C$ & $I_C$ & \emph{sum}\\
\hline
Kohoutek & 08.1977\,-\,11.1979 & 81 & 82 & 100& 0 & 0 & 263 \\
Jurcsik et al. & 07.1990\,-\,07.1993 & 392 & 401 & 403 & 403 &402& 2001 \\
Brno & 08.1999\,-\,11.2002 & 0 &  0 &  196 &201& 159& 556\\
Vy\v{s}kov &05.2001\,-\,09.2002 & 0 &  0 &  267& 283& 0 & 550\\
\hline
 &     $N_{tot}$ &    473 & 483& 966& 887& 561& {\bf 3370}\\
\hline
 &  \emph{obs. error} [mag] & 0.12 &   0.04 &   0.04  &  0.03  &  0.05  &0.06\\
 &  \emph{scatter} [mag] & 0.16 & 0.09 & 0.11 & 0.10 & 0.13 & 0.12\\
 &  \emph{eff. ampl.} [mag] &   0.84  &  0.70 &   0.71 &   0.73&  0.81\\
 \hline
\end{tabular}
\caption[]{The review of photometric observations used: name of
the author or the group, the interval of observations, numbers of
measurements in \emph{UBV}$\!(\!RI)_{\rm{C}}$.} \label{table1}
\end{table}

\begin{figure}
\centerline{\includegraphics[width=0.95\textwidth]{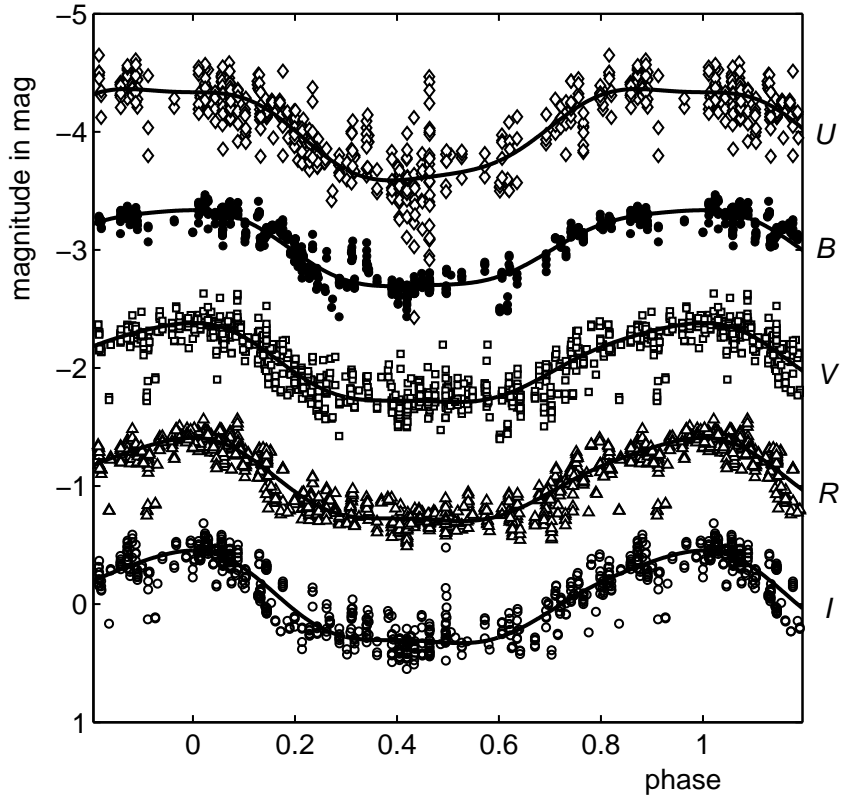}}
\caption[]{The phase diagram of light variations of Sh 2-71 in
\emph{UBV}$\!(\!RI)_{\rm{C}}$ colours for the period $P = 68.132$
d. Plots of individual light curves are mutually shifted by
constant to prevent their interference. In \emph{U} and in lesser
extent in \emph{I} you can see signs of a specific type of the
stellar activity. The observed scatter is significantly larger
than observational errors (see Table I).}
\end{figure}

\section{Ephemeris of light variations}

Analysis of photometric data confirmed the presence of strong more
or less periodic variations of the object in all photometric
colours studied running with the period of about 68 days. 966
individual \emph{V} observations of the star span 135 cycles of
light variations which is quite sufficient for the adequately
accurate determination of linear ephemeris. Using own version of
the non-linear robust regression \cite{Mikulaseketal2003} we have
arrived at the following expression for the moment of the light
maximum:
$$JD_{max}=(2~449~795.85\pm0.22)+(68.132\pm0.005)(E-95).$$
The beginning of the epoch counting was chosen so that the maximum
in $E\!=0$ corresponds to the first light maximum preceding the
first photometric observation of the star.

\noindent This linear ephemeris should be considered only as a
first approximation of the situation. There are strong indications
that both the period and the form of light curves are not stable.
Consequently, the above given uncertainties of both light elements
are only formal.

All checked spurious periods conjugated with the one sidereal day
and the tropic year give markedly worse fit than the 68 d period.

\section{The shapes of light curves}

The comparison of light curves in various colours was done by own
method combining principal component analysis and robust
regression. We have found that periodically repeating parts of
light curves in all colours can be satisfactorily well expressed
as a linear combination of only two principal light curves (see
Fig. 2a), where the first one contains the basic features of
stellar variations and the second one hits mutual differences of
individual light curves. The fit by combination of both basic
curves is displayed in Fig. 1. The dependencies of semi-amplitudes
$A_1$ and $A_2$ corresponding to both principal curves on the
effective wavelength are demonstrated on the Fig. 2b.

\begin{figure}
\tabcapfont \centerline{%
\begin{tabular}{c@{\hspace{1pc}}c}
\includegraphics[width=0.395\textwidth]{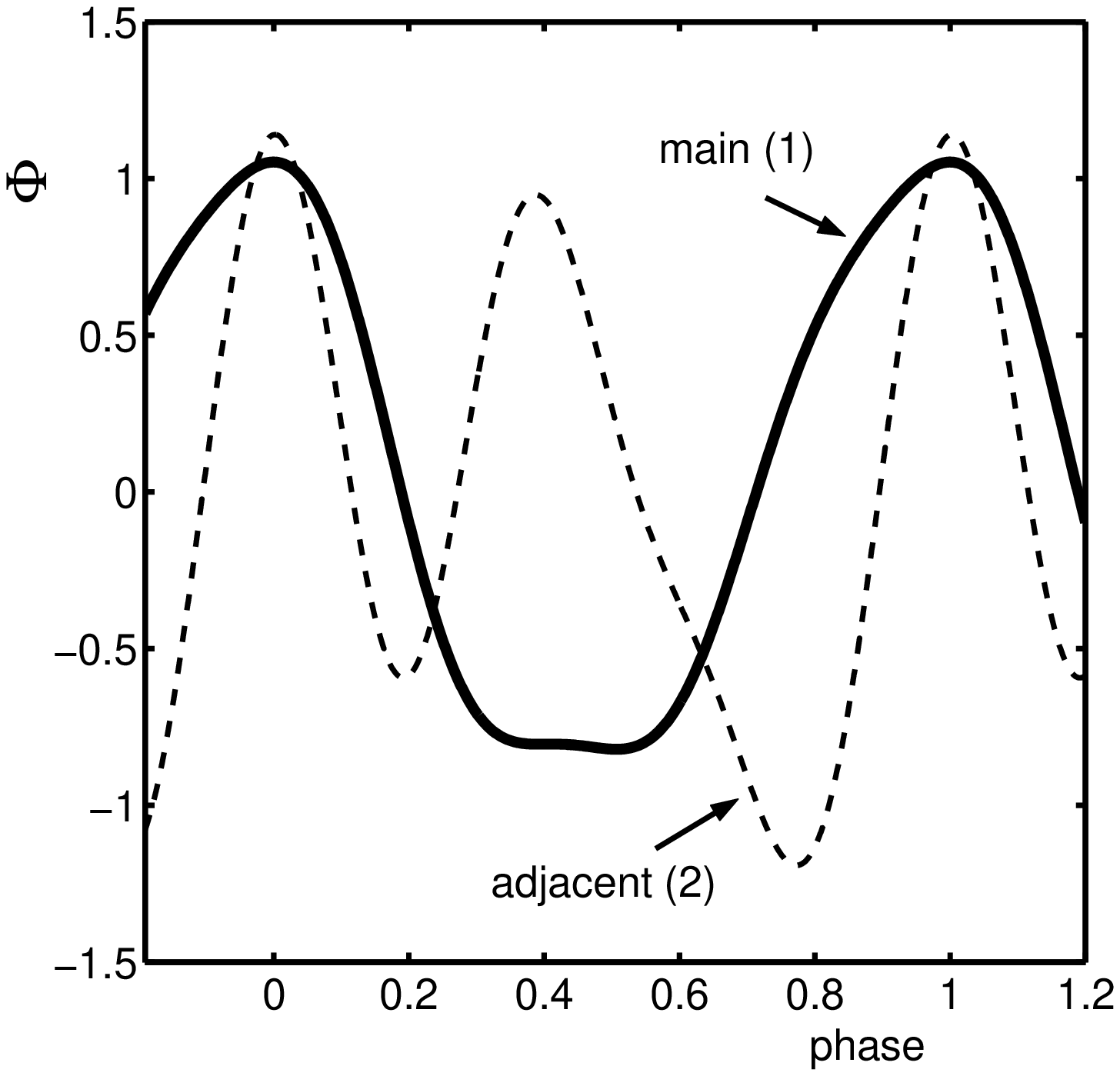}&
\includegraphics[width=0.395\textwidth]{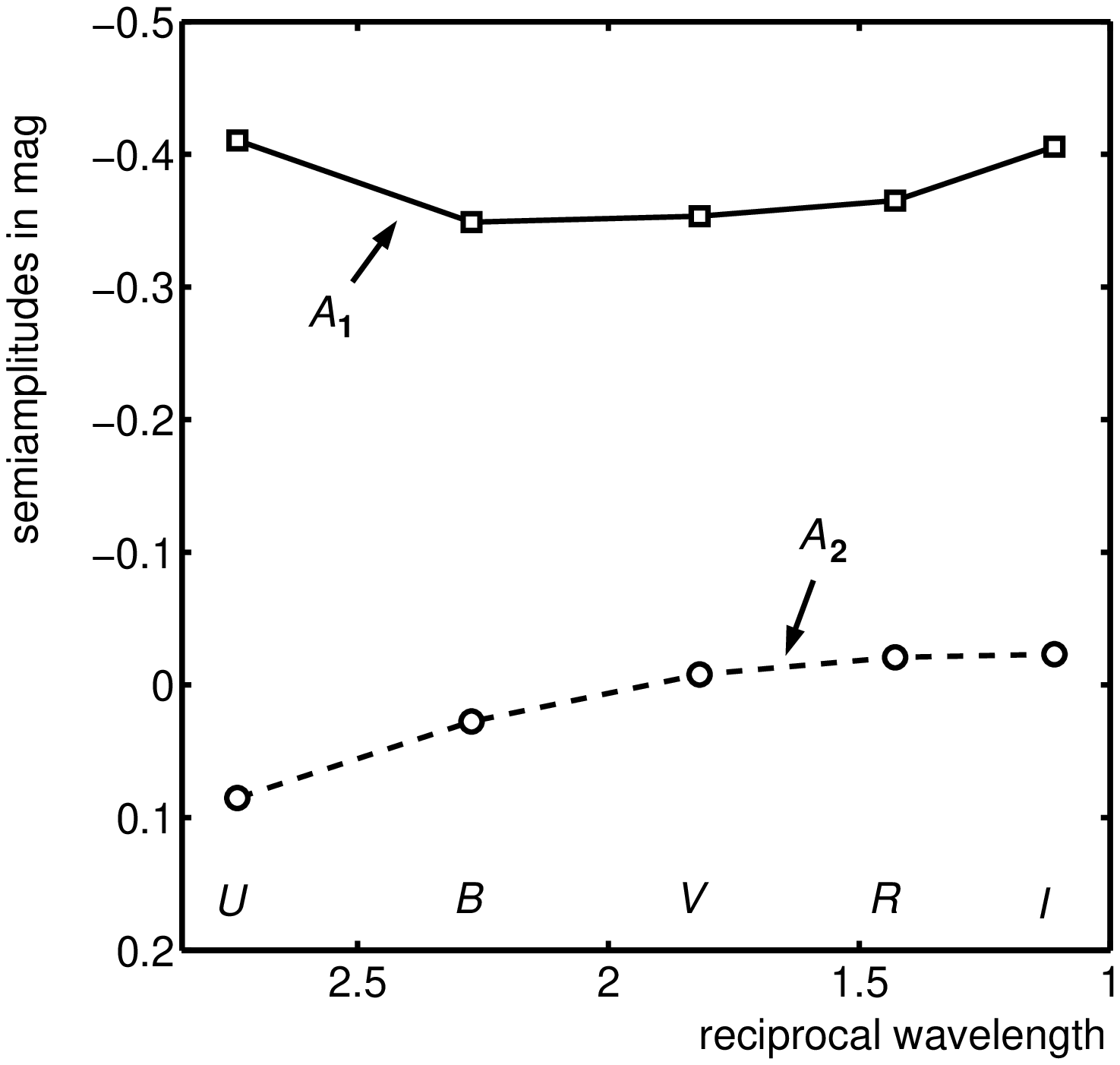}\\
(a) & (b) \\
\end{tabular}}
\caption{(a) The phase graph of the first and second mutually
orthogonal principal functions. (b) The dependence of
semiamplitudes $A_1$ and $A_2$ of the decomposition of light
curves into the first two principal functions on the reciprocal
effective wavelengths expressed in microns.}
\end{figure}

The detailed analysis of photometric behaviour of the object will
be published elsewhere.

\end{article}
\end{document}